\documentclass[journal]{IEEEtran}
\usepackage{amsmath,amsfonts}
\usepackage{algorithmic}
\usepackage{algorithm}
\usepackage{array}
\usepackage[caption=false,font=normalsize,labelfont=sf,textfont=sf]{subfig}
\usepackage{textcomp}
\usepackage{stfloats}
\usepackage{url}
\usepackage{verbatim}
\usepackage{graphicx}
\usepackage{cite}
\usepackage{cleveref}
\usepackage{bm}
\usepackage{amsmath}
\usepackage{empheq}
\usepackage{booktabs}
\usepackage{breqn}
\usepackage{amsfonts,amssymb}
\usepackage[utf8]{inputenc}
\usepackage{xcolor}       % 颜色支持
\usepackage{colortbl}     % 单元格填色
\def\BibTeX{{\rm B\kern-.05em{\sc i\kern-.025em b}\kern-.08em
    T\kern-.1667em\lower.7ex\hbox{E}\kern-.125emX}}
\usepackage{balance}
\begin{document}

\title{Frequency Security Assessment in Power Systems With High Penetration of Renewables Considering Spatio-Temporal Frequency Distribution}

% \author{
% Changjun He,~\IEEEmembership{IEEE Member}, Hua Geng,~\IEEEmembership{IEEE Fellow}, Xiuqiang He,~\IEEEmembership{IEEE Member}, Yushuang Liu,~\IEEEmembership{IEEE Member}
% \thanks{This work was supported by the Fundamental Research Funds for the Central Universities (B250201245) and the National Natural Science Foundation of China (U25B20203). \textit{(Corresponding author: Hua Geng)}}
% \thanks{Changjun He is with the School of Electrical and Power Engineering, Hohai University, Nanjing, 211100 (e-mail: hcj25@hhu.edu.cn).}% <-this % stops a space
% \thanks{Hua Geng and Xiuqiang He are with the Department of Automation, Beijing National Research Center for Information Science and Technology, Tsinghua University, Beijing, 100084, China (e-mail: genghua@tsinghua.edu.cn; xhe@tsinghua.edu.cn).}
% }
\author{
Changjun He,~\IEEEmembership{IEEE Member},Hua Geng,~\IEEEmembership{IEEE Fellow}, Xiuqiang He,~\IEEEmembership{IEEE Member}, Chen Shen,~\IEEEmembership{IEEE Senior Member}, Yushuang Liu,~\IEEEmembership{IEEE Member}
\thanks{This work was supported by the Fundamental Research Funds for the Central Universities (B250201245) and the National Natural Science Foundation of China (U25B20203). \textit{(Corresponding author: XXX)}}
        % <-this % stops a space
\thanks{Changjun He is with the School of Electrical and Power Engineering, Hohai University, Nanjing, 211100 (e-mail: hcj25@hhu.edu.cn).}% <-this % stops a space
\thanks{Hua Geng and Xiuqiang He are with the Department of Automation, Beijing National Research Center for Information Science and Technology, Tsinghua University, Beijing, 100084, China (e-mail: genghua@tsinghua.edu.cn; hxq19@tsinghua.org.cn).}
\thanks{Chen Shen is with the Department of Electrical Engineering, Tsinghua University, Beijing 100084, China (e-mail: shenchen@mail.tsinghua.edu.cn).}
\thanks{Yushuang Liu is with the School of Electrical Engineering and Automation, Wuhan University, Wuhan 430072, China (e-mail: lyshuang@whu.edu.cn).}
}

\markboth{}%
{Shell \MakeLowercase{\textit{et al.}}: A Sample Article Using IEEEtran.cls for IEEE Journals}


\maketitle

\begin{abstract}
The increasing integration of renewable energy sources exacerbates the spatial and temporal differences in frequency across the power system, posing a serious challenge to the accurate and efficient assessment of system frequency security. To address this issue, a generic effective nodal frequency (ENF) model is first established to concisely characterize nodal frequency dynamics. This model is featured by the effective nodal inertia (ENI), damping, and primary regulation parameters, which retain only the dominant constant component governing nodal frequency dynamic performance. This model enables the tractable analytical formulation of nodal frequency trajectory and the key frequency security indicators. Quantitative analysis under the temporary power disturbance condition reveals that the ENI is the most influential parameter governing frequency security. Consequently, the critical nodal inertia for ensuring nodal frequency security is analytically derived. A system-level frequency security index based on the actual ENI and critical nodal inertia is proposed. On the basis of the proposed index, the system frequency security assessment is carried out with the procedure of ``offline calculation and online evaluation'', which is achieved using a lookup table approach and an interpolation method. Simulations on the modified IEEE 39-bus system verify the effectiveness of the proposed assessment method.
\end{abstract}


\section{Introduction}
\IEEEPARstart{W}{ith} the increasing penetration of renewable energy sources (RESs) replacing traditional synchronous generators (SGs) connected to the power system, the inertia and frequency regulation capability of the grid are significantly degraded \cite{11061796}. This seriously endangers the system frequency security (SFS) and even lead to catastrophic blackout events \cite{li2023review}. For example, the UK power grid experienced frequency dip issues and widespread power outages due to insufficient inertia on August 9, 2019 \cite{bialek2020does}. Consequently, system frequency security assessment (SFSA) is imperative for the safe operation of power systems with high RES penetration \cite{ju2021analytic}. SFAS is commonly conducted by checking whether the three main frequency indicators, namely the maximum rate of change of frequency (RoCoF), the maximum frequency deviation, and steady-state frequency, remain within the permissible limits \cite{11304183}. Existing assessment methods are generally categorized into three classes: simulation-based methods, data-based methods, and analytical methods \cite{ju2021analytic}.

Simulation-based methods rely on detailed and comprehensive model of each component and control strategy to simulate system frequency responses under various disturbances. By examining the frequency trajectories, whether the frequency indicators violate the security limits is determined. In reference \cite{8959148}, frequency security in power systems with high penetration of converter-based RESs is analyzed via electromagnetic transients simulations. It is found that an all-converter system exhibits more reliability than the hybrid SG-RES system. However, the offline nature of these methods limit their application in real-time online assessment \cite{9465818}. Besides, if the power disturbance changes, the simulation must be run again.

Data-driven methods leverage measurement data from phasor measurement units (PMUs) to train intelligence models (e.g., graph neural networks) for SFSA \cite{11304183}. While offering high accuracy, these methods often lack physical interpretability and fail to reveal the underlying frequency mechanisms.

In contrast, analytical methods feature clear physical mechanisms and explicit mathematical expressions, making them the most preferred for SFSA \cite{10102594}. The average system frequency (ASF) model and the system frequency response (SFR) model are the two mainstream analytical frameworks. Based on the model, the three frequency security indicators is analytically derived to assess the frequency security. Reference \cite{10102594} established a generic ASF model, considering the different governing systems of the SGs. The maximum frequency deviation is calculated based on the model to determine the frequency security. However, the constraints on the other two indicators — RoCoF and the steady-state frequency — are ignored. An extended ASF model considering the effects of frequency deviations on boiler auxiliaries is developed for SFSA in reference \cite{9130687}, which allows a more accurate SFSA. In reference \cite{zhang2023security}, an SFR model that considers the virtual inertia and fast frequency regulation of RESs is constructed. It is discovered that inertia is significant for SFS and the minimum inertia to guarantee SFS is derived \cite{9347330}. A novel inertia security ratio is defined in \cite{10577465} as an index for SFS. Nevertheless, in power systems with high penetration of RESs, the frequency and inertia feature significant spatiotemporal heterogeneity. The frequency responses at different areas and times are inconsistent \cite{8630092,he2024analysis}. The conventional ASF and SFR models, which base on the assumption of a uniform frequency distribution and only analyze the average system frequency dynamics, are no longer applicable in SFSA. Even if the system average dynamic performance meets the security criteria, localized frequency violations may exceeds the limit and lead to frequency security issues, as shown in Fig.\ref{fig.1}. Reference \cite{11160220} explores the dynamic inertia of each node, considering the spatiotemporal heterogeneity of inertia distribution. An SFSA index base on dynamic inertia contribution is developed. However, the performance of the three frequency security indicators is unclear.

To fill the research gap, this paper developed a frequency security index for online SFSA based on the nodal inertia, accounting for the spatiotemporal distribution of frequency dynamics. The main contributions are summarized as follows:
\begin{enumerate}
    \renewcommand{\labelenumi}{(\arabic{enumi})}
    \item \textbf{Frequency modeling}: To characterize the frequency performance of individual nodes while circumventing the complexity of time-varying parameters, a generic effective nodal frequency (ENF) model is established. This model simplifies the nodal frequency dynamics using constant effective parameters, including effective nodal inertia (ENI), damping, and primary regulation.
    \item \textbf{Frequency mechanism}: Based on the ENF model, analytical expressions are derived for nodal frequency trajectories and the three frequency security indicators. Quantitative analysis shows that the ENI has the greatest influence on nodal frequency security performance compared to other parameters.
    \item \textbf{Security index}: The critical nodal inertia required to maintain frequency security is analytically calculated. A system-level frequency security index is then constructed based on the actual ENI and critical nodal inertia.
    \item \textbf{Assessment method}: An ``offline calculation and online evaluation" strategy is designed using a lookup table approach, enabling accurate and rapid SFSA that considers the spatiotemporal frequency distribution.
\end{enumerate}

Simulations are conducted on the modified IEEE 39-bus system to verify the effectiveness of the proposed method.

\section{Simplified Modeling of Frequency Distribution}
\begin{figure}[t]
\centering
\includegraphics[width=88mm]{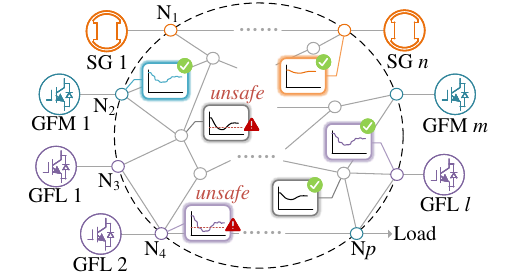}
\caption{Topology and frequency distribution of a power system integrated with high penetration of RESs.}
\label{fig.1}
\end{figure}

A power system with high penetration of RESs is shown in Fig.\ref{fig.1}. The generators contain traditional SGs, grid-forming (GFM) inverters, and grid-following (GFL) inverters. Suppose there are $n$ SGs, $m$ GFMs, and $l$ GFLs, with $p$ main nodes.

\subsection{Time-varying Nodal Frequency Model}
Basically, the frequency dynamics of all SGs, GFL devices, and GFM devices can be converter into the form as the SFR model \cite{10844898}. Consequently, all nodes’ frequency dynamics can also be represented as the SFR model as below.

\begin{subequations}\label{eq:exact nodal frequency}
\begin{align}
2{H_{j}}\left( t \right)\frac{{{{d}}{\omega _j}\left( t \right)}}{{{{d}}t}} & = \Delta {P_i} - {D_{j}}\left( t \right)\left( {{\omega _j}\left( t \right) - {\omega _{\rm{0}}}} \right) - g\left( t \right) \label{eq:exact nodal frequency a} \\ 
{\tau _{j}}\left( t \right)\frac{{{{d}}g\left( t \right)}}{{{{d}}t}} & = {K_{j}}\left( t \right)\left( {{\omega _j}\left( t \right) - {\omega _{\rm{0}}}} \right) - g\left( t \right) \label{eq:exact nodal frequency b}
\end{align}
\end{subequations}
where $t$ represents the time variable. $\Delta P_{i}$ represents the power disturbance at node $i$. This also presents the imbalance between the power generation and consumption. $\omega_j(t)$ is the frequency of node $j$. $g(t)$ is the equivalent primary regulation power of the node $j$. $H_{j}(t)$, $D_{j}(t)$, $K_{j}(t)$, and $\tau_{j}(t)$ are the equivalent inertia, damping, primary regulation, and time-delay of node $j$ when there is a power disturbance at node $i$, respectively. These parameters for the node $j$ are dependent on those parameters of the generators (SGs, GFLs, and GFMs), as well as their capability combinations and the system structures. Considering that the corresponding parameters of the GFL and GFM are non-constant under the influence of multi-timescale control strategies, as well as the volatility of renewables, $H_{j}(t)$, $D_{j}(t)$, $K_{j}(t)$, and $\tau_{j}(t)$ are also time-varying. The time-varying nodal inertia ($H_{j}(t)$) and the frequency dynamics ($\omega_j(t)$) of the node $j$ are drawn in the blue box in Fig. \ref{fig.2}. It is evident that the effective nodal inertia and the associated nodal frequency both exhibit complex behaviors.

\subsection{Proposed Effective Nodal Frequency Model}
\begin{figure*}[t]
\centering
\includegraphics[width=180mm]{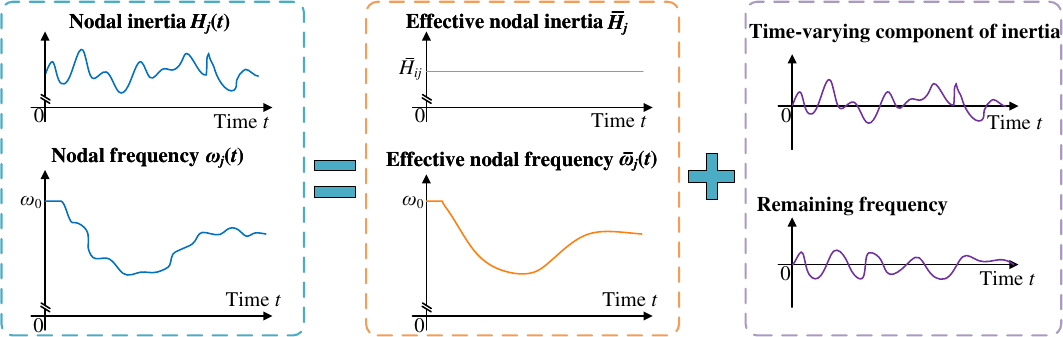}
\caption{The actual nodal inertia/frequency, the effective nodal inertia/frequency, and the remaining  components.}
\label{fig.2}
\end{figure*}

Regarding that the time-varying model is difficult to analyze, the ENI, the effective nodal damping, the effective nodal primary regulation, and effective time delay are defined. These parameters are constant components of the actual time-varying parameters and play the most important role in frequency dynamic performance. As the same, the ENF is defined corresponding to the effective parameters.

\begin{subequations}\label{eq:exact parameters}
\begin{align}
{H_{j}}\left( t \right) &= {{\bar H}_{j}} + h_j(t) \label{eq:H}\\
{D_{j}}\left( t \right) &= {{\bar D}_{j}} + d_j(t) \label{eq:D}\\
{K_{j}}\left( t \right) &= {{\bar K}_{j}} + k_j(t) \label{eq:K}\\
{\tau _{j}}\left( t \right) &= {{\bar \tau }_{j}} + e_j(t) \label{eq:tau}\\
\omega_j(t) &= \bar\omega_j(t) + w_j(t)
\end{align}
\end{subequations}
where ${\bar H_{j}}$, $\bar D_{j}$, $\bar K_{j}$, and $\bar \tau_{j}$ are the constant effective frequency regulation parameters of the node $j$. ${{h_{j}(t)}}$, ${{d_{j}(t)}}$, ${{k_{j}(t)}}$, and ${{e_{j}(t)}}$ are the remaining time-varying components. $\bar\omega_j(t)$ is the ENF. $w_j(t)$ is the remaining high-frequency component of the nodal frequency trajectory. The relationship between the effective nodal parameters and the ENF is called the effective nodal frequency model.
\begin{subequations}\label{eq:simple nodal frequency}
\begin{align}
2{{\bar H_{j}}}\frac{{d{{\bar \omega }_j}\left( t \right)}}{{dt}} &= \Delta {P_i} - {{\bar D_{j}}}\left( {{{\bar \omega }_j}\left( t \right) - {\omega _{\rm{0}}}} \right) - {{\bar g}_j}\left( t \right) \label{eq:simple nodal frequency 1}\\
{{\bar \tau_{j} }}\frac{{d{{\bar g}_j}\left( t \right)}}{{dt}} &= {{\bar K_{j}}}\left( {{{\bar \omega }_j}\left( t \right) - {\omega _{\rm{0}}}} \right) - {{\bar g}_j}\left( t \right) \label{eq:simple nodal frequency 2}
\end{align}
\end{subequations}
where ${\bar g_j}$ is donated as the effective primary regulation power of the node $j$. In Fig.\ref{fig.2}, the actual nodal inertia/frequency, the effective nodal inertia/frequency, and the remaining sine/cosine component of the inertia/frequency are drawn. It can be seen that the effective nodal frequency retains the core dynamics characteristics that play the most important role in system frequency security. The ENF model could balance conciseness and accuracy, which enables conducting analytical calculations. Besides, the frequency regulation parameters, such as the ENI, are constant and suitable for engineering practice.

\section{Data-Based Analysis of Nodal Frequency}
This section studies the analytical expressions of the ENF model using the Laplace transform method and the node-specific frequency trajectory base on the power/frequency data.

\subsection{Unified Analytical Solution for ENF Model}
The power disturbances can be divided into two categories when considering frequency stability: permanent and temporary power disturbances. The former usually contains a sudden cut off of loads or generators. The latter typically occurs when there is a ground fault or a temporary circuit interruption. 

\subsubsection{Permanent Power Disturbance Scenarios} When the load at node $i$  is increased by $\Delta P$, or when the power generation at node $i$ suddenly decreases by $\Delta P$. there will be a power disturbance of $\Delta P_i = -\Delta P$. Accordingly, the solution for the ENF model \eqref{eq:simple nodal frequency} is derived using the Laplace transforms method as \eqref{eq:frequency_per_solu} \cite{leon2024frequency}. More details are in Appendix.
\begin{equation}
\label{eq:frequency_per_solu}
\bar \omega_j(t) = \omega_0 + \frac{\Delta P_i}{\bar D_{j}+\bar K_{j}} \left( 1 - e^{-\lambda t} \left( \cos\omega_d t - A_1 \sin\omega_d t \right) \right)
\end{equation}
where $\lambda$, $\omega_d$, and $A_1$ are calculated in Appendix.

\subsubsection{Temporary Power Disturbances Scenarios} 
\begin{figure}[!b]
\centering
\includegraphics[width=78mm]{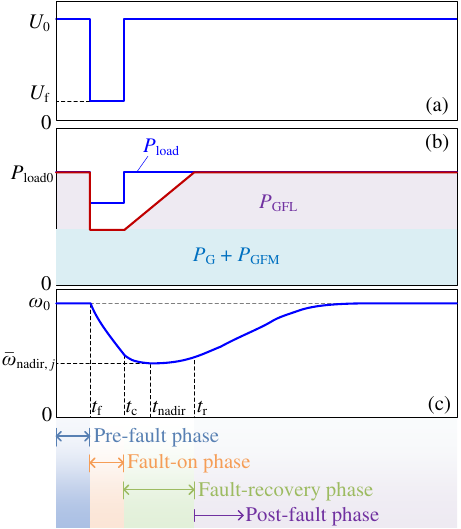}
\caption{Transient process under the temporary power disturbances scenario. (a) Grid voltage. (b) Active power. (c) Nodal Frequency}
\label{fig.3}
\end{figure}

Some disturbances, such as a grid fault, will arouse a temporary power disturbance. The transient process is more complicated, which can be divided into four stages, as shown in Fig.\ref{fig.3}.
\begin{enumerate}
    \renewcommand{\labelenumi}{(\arabic{enumi})}
    \item Pre-fault phase ($0 \leq t \leq t_{\rm{f}}$). The nodal frequency runs stably at $\bar \omega_j(t)= \omega_0$ during the pre-fault stage.
    \item Fault-on phase  ($t_{\rm{f}} < t \leq t_{\rm{c}} $). The fault occurs at $t_{\rm{f}}$ and clears at $t_{\rm{c}}$. During this stage, the voltage dip leads to a power consumption decrease on the loads of $\Delta P_{\rm{load}} = (1-aU^2_{\rm{f}}-bU_{\rm{f}}-c)P_{\rm{load0}}$. $U_{\rm{f}}$ is the fault voltage and $P_{\rm{load0}}$ is the nominal load power consumption. $a$, $b$, $c$ are the proportions of the resistive load, current load and power load, respectively. $a+b+c=1$.  Besides, $i_d = 0   (P_{\rm{GFL}} = 0)$ is the usually used strategy for GFLs during grid faults, resulting in a power generation reduction of $\Delta P_{\rm{gene}} = -P_{\rm{GFL0}}$. $P_{\rm{GFL0}}$ is the active power of all GFLs before the fault. The mechanical power of SGs and the power reference of GFMs remain unchanged. As a result, the power disturbance during the fault-on stage is a constant value of $\Delta P_{i2} = \Delta P_{\rm{gene}} + \Delta P_{\rm{load}} = -P_{\rm{GFL0}} +(1-aU^2_{\rm{f}}-bU_{\rm{f}}-c)P_{\rm{load0}}$.
    \item Fault-recovery phase ($t_{\rm{c}} < t \leq t_{\rm{r}}$). After the fault is cleared at time $t_{\rm{c}}$, power consumption of the load soon recovers back to the pre-fault value of $P_{\rm{load0}}$. The active power of the GFL recovers at a ramp of $r_{\rm{p}}$  (pu/s). Therefore, the power imbalance is $\Delta P_{i3} = [r_{\rm{p}}(t-t_{\rm{c}}) – 1]P_{\rm{GFL0}}$ at this stage.
    \item Post-fault phase ($t > t_{\rm{r}}$). The power imbalance is equal to 0 at $t_{\rm{r}} = t_{\rm{c}}+1/r_{\rm{p}})$, i.e. $\Delta P_{i4} = 0$. The nodal frequency finally recovers to the rated value through the action of the inertia, damping, and primary regulation.
\end{enumerate}

\begin{figure*}[t]
  \centering
  \begin{equation}
\label{eq:ENF2}
{\bar \omega _j}\left( t \right) = \left\{ \begin{array}{l}
{\omega _{\rm{0}}}, 0 \le t \le {t_{\rm{f}}}\\

\omega_0 + K_2\left( 1 - {e^{ - \lambda \left( {t - {t_{\rm{f}}}} \right)}} 
{\left( {\cos \left( {{\omega _d}\left( {t - {t_{\rm{f}}}} \right)} \right) - A_1\sin \left( {{\omega _d}\left( {t - {t_{\rm{f}}}} \right)} \right)} \right)}
 \right), t_{\rm{f}}<t\le t_{\rm{c}}\\

{\omega _0} + {C_1}\left( {t - {t_{\rm{c}}}} \right) + {C_2} + {e^{ - \lambda \left(t-t_{\rm{c}}\right)}}  {\left( {{A_{\cos }}\cos \left( {{\omega _d}\left( {t - {t_{\rm{c}}}} \right)} \right) + {A_{\sin }}\sin \left( {{\omega _d}\left( {t - {t_{\rm{c}}}} \right)} \right)} \right)}, t_{\rm{c}}<t\le t_{\rm{r}}\\

{\omega _0} + {e^{ - \lambda (t - {t_r})}} \left({{{\bar \omega }_j}\left( {{t_{\rm{r}}}} \right)\cos ({\omega _d}(t - {t_r})) + }{\frac{{{{\bar \omega '}_j}({t_r}) + \lambda {{\bar \omega }_j}\left( {{t_{\rm{r}}}} \right)}}{{{\omega _d}}}\sin ({\omega _d}(t - {t_r}))}\right)
 , t>t_{\rm{r}}
\end{array} \right.
\end{equation}
  %\caption{expression}
\end{figure*}

Accordingly, the whole fault process is plotted in Fig.\ref{fig.3}. The analytical time-domain solution under the temporary power disturbance condition is derived using the Laplace method as \eqref{eq:ENF2}. More details are shown in Appendix.

\subsection{Data-Based Calculation for Nodal Frequency}
\begin{figure}[!t]
\centering
\includegraphics[width=88mm]{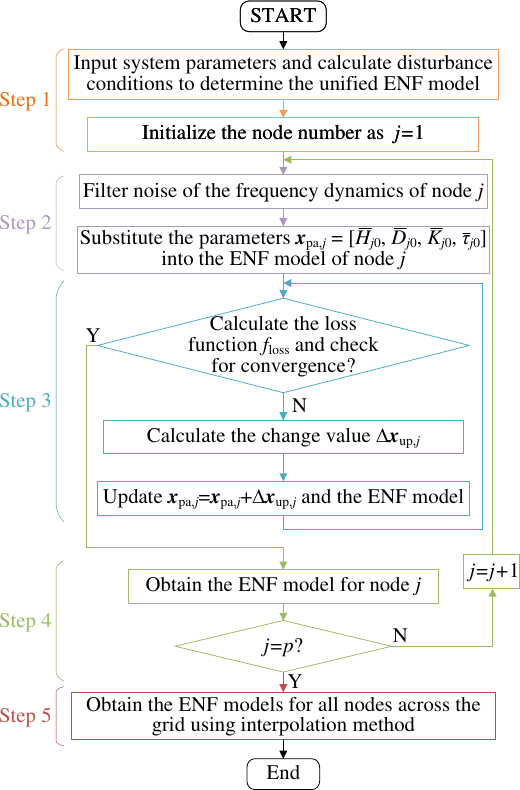}
\caption{Flow of data-based calculation for nodal frequency.}
\label{fig.4}
\end{figure}

To obtain the frequency distribution across the grid. The specific expression of the ENF dynamics of each node is required. A data-based dynamics fitting approach based on the simplex method is proposed to obtain the frequency dynamics of each node, along with the ENI, damping, and primary frequency regulation parameters. The flow can be divided into 5 steps, as plotted in Fig.\ref{fig.4}.

\subsubsection{Step 1: Data Acquisition and Node Number Initialization} First, input the system parameters (such as node numbers, generator numbers/types) and the disturbance conditions (disturbance type, location node ${\rm{N}}_i$, and magnitude). Using the PMUs to acquire the frequency dynamics data of each node. Initialize the node number as $j = 1$.

\subsubsection{Step 2: ENF Model Initialization} Filter the frequency response curve of node $j$ to eliminate the measurement noise. Then substitute the initial frequency regulation parameters of ${\textbf{x}_{{\rm{pa}},j}} = \left[ {{{\bar H}_{j0}},{{\bar D}_{j0}},{{\bar K}_{j0}},{{\bar \tau }_{j0}}} \right]$ into \eqref{eq:ENF2} to determine the initial ENF model of the node $j$.

\subsubsection{Step 3: Parameters Update} Calculate the loss function $f_{\rm{loss}}$ of the node $j$, which represents the error between the actual nodal frequency and the ENF (the expression of $f_{\rm{loss}}$ will be explained later). If the loss function converges, the ENF model of the node $N_j$ is obtained and go to Step 4. Otherwise, if the loss function is not converged, calculate the change value of the parameters with $\Delta {\textbf{x}_{{\rm{up}},j}} = \left[ {\Delta {{\bar H}_{j}},\Delta {{\bar D}_{j}},\Delta {{\bar K}_{j}},\Delta {{\bar \tau }_{j}}} \right]$ using the simplex method. The updated frequency regulation parameters vector is iterated as $\textbf{x}_{{\rm{pa}},j} = \textbf{x}_{{\rm{pa}},j} + \Delta \textbf{x}_{{\rm{up}},j}$. Then, update the ENF model and return to the start of Step 3.

The loss function $f_{\rm{loss}}$: When conducting the frequency security assessment, the main focus is on whether the three critical indicators, i.e. RoCoF (denoted as $R_{\max}$), the maximum frequency deviation (denoted as $\omega_{\rm{nadir}}$), and the steady-state frequency (denoted as $\omega_{\rm{steady}}$), exceed the limits. Consequently, in addition to the alignment between the ENF dynamics and the actual node frequency dynamics, the accuracy of these three key indicators is also of great importance. The ENF dynamics must not only closely match the actual nodal frequency dynamics but also capture these indicators. The maximum RoCoF usually occurs at the moment of the disturbance. In engineering practice, the RoCoF is typically calculated by taking the frequencies at two adjacent time points with a small time interval. Therefore, the frequencies of $\omega_j(t_{\rm{f}})$ at time $t_{\rm{f}}$ and $\omega_j(t_{\rm{RoCoF}})$ at $t_{\rm{RoCoF}}$ ($t_{\rm{RoCoF}} - t_{\rm{f}}$ is about 100--200 ms) determine the maximum RoCoF as $R_{\max} = (\omega_j(t_{\rm{RoCoF}}) - \omega_j(t_{\rm{f}}))/(t_{\rm{RoCoF}} - t_{\rm{f}})$. Since the frequency at $t_{\rm{f}}$ is equal to the nominal frequency, $R_{\max}$ is only determined by the accuracy of $\omega_j(t_{\rm{RoCoF}})$. Based on this, the loss function $f_{\rm{loss}}$ is proposed as follows.

\begin{equation}\label{eq:loss function}
\begin{split}
{f_{{\rm{loss}}}} = &{k_1}\frac{{\int_{{t_{\rm{f}}}}^{{t_1}} {\left( {{{\left\| {{{\bar \omega }_j}\left( t \right) - {\omega _j}\left( t \right)} \right\|}^2}} \right){\rm{d}}t} }}{{{t_1} - {t_{\rm{f}}}}} \\
& + {k_2}{\left\| {{{\bar \omega }_j}\left( {{t_{\rm{RoCoF}}}} \right) - {\omega _j}\left( {{t_{\rm{RoCoF}}}} \right)} \right\|^2} \\
& + {k_3}{\left\| {\arg \mathop {\min }\limits_t {\{ {{\bar \omega }_j}\left( t \right) \}} - \arg \mathop {\min }\limits_t \{ {{\omega _j}\left( t \right)} \} } \right\|^2} \\
& + {k_4}{\left\| {\min \{ {{{\bar \omega }_j}\left( t \right)} \} - \min \{ {{\omega _j}\left( t \right)} \}} \right\|^2} \\
& + {k_5}{\left\| {{{\bar \omega }_j}\left( {{t_\infty }} \right) - {\omega _j}\left( {{t_\infty }} \right)} \right\|^2}
\end{split}
\end{equation}
where $t_1$ denotes the end time of the curve fitting. $k_1, k_2, k_3, k_4$, and $k_5$ are the fitting coefficients for frequency dynamics, maximum RoCoF, time of the maximum frequency deviation, the maximum frequency deviation, and steady-state frequency, respectively.

\subsubsection{Step 4: Node Number Update} If $j$ is equal to $p$, skip to Step 5. Otherwise, if $j$ is not equal to $p$, $j = j + 1$ and return to the start of Step 2.

\subsubsection{Step 5: Supplement of Frequency Distribution} Supposing the node $N_m$ is directly connected near to a set of nodes denoted as $Q$, then the frequency regulation parameters of the nodal $N_m$ ($\textbf{x}_{{\rm{pa}},m} = \left[ {{\bar H}_{\rm{m}}},{{\bar D}_{\rm{m}}},{\bar K}_{\rm{m}},{{\bar \tau }_{\rm{m}}} \right] $) can be calculated using the interpolation method, as \eqref{eq:interpolation method} shown. 
\begin{equation}\label{eq:interpolation method}
{\textbf{x}_{{\rm{pa,}}m}} = \frac{{\sum\limits_{q \in Q} \frac{\textbf{x}_{{\rm{pa,}}q}}{{l_{mq}}} }}{{\sum\limits_{q \in Q} \frac{1}{{l_{mq}}} }}
\end{equation}
where $\textbf{x}_{{\rm{pa}},q} =  \left[ {{{\bar H}_{q}},{{\bar D}_{q}},{{\bar K}_{q}},{{\bar \tau }_{q}}} \right]$ is the frequency regulation parameters vector for the node ${\rm{N}}_q$. $l_{mq}$ is the transmission line length between the nodes ${\rm{N}}_m$ and ${\rm{N}}_q$. Finally, the ENF model of any node is determined based on \eqref{eq:simple nodal frequency}.

\subsection{Validation of ENF Model Accuracy}
\begin{figure}[t]
\centering
\includegraphics[width=88mm]{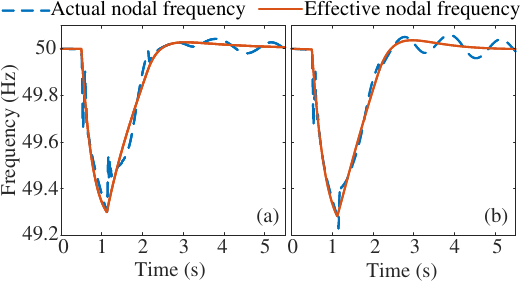}
\caption{The actual nodal frequency and ENF at nodes (a) ${\rm{N}}_1$ and (b) ${\rm{N}}_{16}$.}
\label{fig.5}
\end{figure}

Simulations are conducted in the modified IEEE 39-bus system with high penetration of RESs, with the topology shown in Fig.7 and parameters depicted in TABLE II in Section V. Analysis above shows that the temporary power disturbance conditions are more complicated than the permanent power disturbance conditions. Hereafter, the temporary power disturbance condition is studied as an example for analysis. The frequencies of the nodes ${\rm{N}}_1$ and ${\rm{N}}_{16}$ are plotted in Fig.\ref{fig.5} (a) and (b), respectively. The actual nodal frequency and the ENF trajectories are drawn with red solid lines and blue dashed lines, respectively. It implies that the actual nodal frequency exhibits complex dynamics with superimposed high-frequency components. The ENF is simpler while retaining the key frequency security indicators and frequency dynamics. The overall trend is effectively captured by the ENF trajectory. TABLE \ref{tab:freq_error} show the errors (equation \eqref{eq:ENF2}) of all the 39 nodes, with the maximum error occurs at node ${\rm{N}}_{38}$ of 2.29 \%. This confirm that the ENF accurately captures both the transient frequency dynamics and key frequency security indicators. 

\begin{table}[t]
  \centering
  % 调整行高，彻底消除单元格上下留白（数值1.2为最佳，可微调）
  \renewcommand{\arraystretch}{1.2}
  \caption{Errors Between Actual Nodal Frequency and ENF}
  \begin{tabular}{l *{8}{c}}
    \hline
    Node    & ${\rm{N}}_1$    & ${\rm{N}}_2$    & ${\rm{N}}_3$    & ${\rm{N}}_4$    & ${\rm{N}}_5$    & ${\rm{N}}_6$    & ${\rm{N}}_7$    & ${\rm{N}}_8$ \\
    Error (\%) & \cellcolor{blue!13}1.85 & \cellcolor{blue!12}1.84 & \cellcolor{blue!12}1.84 & \cellcolor{blue!12}1.84 & \cellcolor{blue!13}1.85 & \cellcolor{blue!13}1.85 & \cellcolor{blue!13}1.85 & \cellcolor{blue!13}1.85 \\
    \hline
    Node    & ${\rm{N}}_9$    & ${\rm{N}}_{10}$   &${\rm{N}}_{11}$   & ${\rm{N}}_{12}$   & ${\rm{N}}_{13}$   & ${\rm{N}}_{14}$   & ${\rm{N}}_{15}$   & ${\rm{N}}_{16}$ \\
    Error (\%) & \cellcolor{blue!13}1.85 & \cellcolor{blue!12}1.84 & \cellcolor{blue!12}1.84 & \cellcolor{blue!12}1.84 & \cellcolor{blue!12}1.84 & \cellcolor{blue!12}1.84 & \cellcolor{blue!12}1.83 & \cellcolor{blue!12}1.83 \\
    \hline
    Node    & ${\rm{N}}_{17}$   & ${\rm{N}}_{18}$   & ${\rm{N}}_{19}$   & ${\rm{N}}_{20}$   & ${\rm{N}}_{21}$   & ${\rm{N}}_{22}$   & ${\rm{N}}_{23}$   & ${\rm{N}}_{24}$ \\
    Error (\%) & \cellcolor{blue!12}1.84 & \cellcolor{blue!12}1.84 & \cellcolor{blue!11}1.81 & \cellcolor{blue!10}1.80 & \cellcolor{blue!11}1.82 & \cellcolor{blue!11}1.82 & \cellcolor{blue!11}1.82 & \cellcolor{blue!11}1.82 \\
    \hline
    Node    & ${\rm{N}}_{25}$   & ${\rm{N}}_{26}$   & ${\rm{N}}_{27}$   & ${\rm{N}}_{28}$   & ${\rm{N}}_{29}$   & ${\rm{N}}_{30}$   & ${\rm{N}}_{31}$   & ${\rm{N}}_{32}$ \\
    Error (\%) & \cellcolor{blue!12}1.84 & \cellcolor{blue!11}1.82 & \cellcolor{blue!12}1.83 & \cellcolor{blue!21}1.98 & \cellcolor{blue!27}2.08 & \cellcolor{blue!13}1.85 & \cellcolor{blue!13}1.85 & \cellcolor{blue!12}1.83 \\
    \hline
    Node    & ${\rm{N}}_{33}$   & ${\rm{N}}_{34}$   & ${\rm{N}}_{35}$   & ${\rm{N}}_{36}$   & ${\rm{N}}_{37}$   & ${\rm{N}}_{38}$   & ${\rm{N}}_{39}$   & \\
    Error (\%) & \cellcolor{blue!11}1.81 & \cellcolor{blue!14}1.87 & \cellcolor{blue!12}1.83 & \cellcolor{blue!11}1.82 & \cellcolor{blue!12}1.84 & \cellcolor{blue!40}2.29 & \cellcolor{blue!13}1.85 & \\
    \hline
  \end{tabular}
  \label{tab:freq_error}
\end{table}

\section{System Frequency Security Assessment Based on Nodal Inertia}
\subsection{Analysis on Key Frequency Security Indicators}
As is known, the frequency security is assessed based on three main indicators: maximum ROCOF, the maximum frequency deviation, and steady-state frequency.

\subsubsection{Maximum RoCoF} Based on \eqref{eq:ENF2}, the maximum RoCoF of the node $j$ is derived  as follows.
\begin{equation}\label{ROCOF}
{R_{\max ,j}} \approx {\bar R_{\max ,j}} = \frac{{\Delta {P_{i2}}}}{{{{\bar H}_{j}}}}
\end{equation}

\subsubsection{Maximum Frequency Deviation} Take the frequency drop case as an example, the maximum frequency deviation can be calculated by taking the smallest of the four stage minimums. Ignore the error between the actual nodal frequency and the ENF, the maximum frequency deviation of the node $j$ is calculated as follows.
\begin{equation}\label{frequency nadir}
{\omega _{{\rm{nadir}},j}} \approx {\bar \omega _{{\rm{nadir}},j}} = \min \left\{ {{\omega _0},{{\bar \omega }_j}\left( {{t_{{\rm{m2}}}}} \right),{{\bar \omega }_j}\left( {{t_{{\rm{m3}}}}} \right),{{\bar \omega }_j}\left( {{t_{{\rm{m4}}}}} \right)} \right\}
\end{equation}
where $t_{\rm{m2}}, t_{\rm{m3}}$, and $t_{\rm{m4}}$ are the times of the minimum frequency in fault-on, fault-recovery, and post-fault phases, respectively. Their analytical expression are shown in Appendix B. The analytical expression of the frequency nadir point can also be found in Appendix.

\subsubsection{Steady-State Frequency} After a temporary power disturbance, both of the generation power and the load consumption will recover to the pre-fault values. Consequently, the steady-state frequency is equal to the nominal frequency.

\subsection{Sensitivity Analysis of Frequency Regulation Parameters} 
\begin{figure}[b]
\centering
\includegraphics[width=88mm]{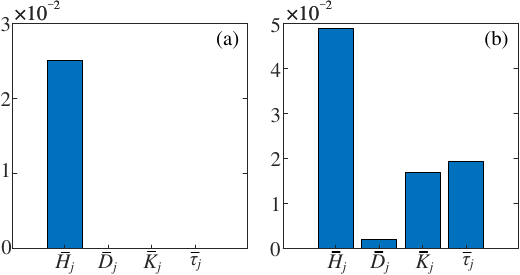}
\caption{Unit-less sensitivity of (a) the maximum RoCoF and (b) the maximum frequency deviation to the four frequency regulation parameters.}
\label{fig.7}
\end{figure}

The four frequency regulation parameters govern the nodal frequency dynamics and the frequency security indicators. The unit-less sensitivity values of the maximum RoCoF to the four parameters are derived as \eqref{eq:S} based on \eqref{ROCOF}.
\begin{equation}\label{eq:S}
\textbf{S}_{{\rm{R}},j} = \left[ {\begin{array}{*{20}{c}}
{\frac{{\partial {R_{\max ,j}}}}{{\partial {{\bar H}_{j}}}} \cdot \frac{{{{\bar H}_{j}}}}{{{R_{\max ,j}}}}}\\
{\frac{{\partial {R_{\max ,j}}}}{{\partial {{\bar D}_{j}}}} \cdot \frac{{{{\bar D}_{j}}}}{{{R_{\max ,j}}}}}\\
{\frac{{\partial {R_{\max ,j}}}}{{\partial {{\bar K}_{j}}}} \cdot \frac{{{{\bar K}_{j}}}}{{{R_{\max ,j}}}}}\\
{\frac{{\partial {R_{\max ,j}}}}{{\partial {{\bar \tau }_{j}}}} \cdot \frac{{{{\bar \tau }_{j}}}}{{{R_{\max ,j}}}}}
\end{array}} \right] = \left[ {\begin{array}{*{20}{c}}
{ - \frac{{\Delta {P_{i2}}}}{{{{\bar H}_{j}}{\bar R_{\max ,j}}}}}\\
0\\
0\\
0
\end{array}} \right]
\end{equation}
where $\textbf{S}_{{\rm{R}},j}$ is the sensitivity matrix of the maximum RoCoF to the regulation parameters. The result is plotted in Fig.\ref{fig.7}(a). It can be seen that the sensitivity of the nodal inertia relative to the maximum RoCoF is the largest. While the sensitivity of other parameters is equal to zero.

Similarly, the unit-less sensitivity matrix of the maximum frequency deviation to the parameters is derived baed on \eqref{frequency nadir}.
\begin{equation}\label{eq:Smin}
\textbf{S}_{{\rm{nadir}},j} = \left[ {\begin{array}{*{20}{c}}
{\frac{{\partial {{\omega }_{{\rm{nadir}},j}}}}{{\partial {{\bar H}_{j}}}} \cdot \frac{{{{\bar H}_{j}}}}{{{{\omega }_{{\rm{nadir}},j}}}}}\\
{\frac{{\partial {{\omega }_{{\rm{nadir}},j}}}}{{\partial {{\bar D}_{j}}}} \cdot \frac{{{{\bar D}_{j}}}}{{{{\omega }_{{\rm{nadir}},j}}}}}\\
{\frac{{\partial {{\omega }_{{\rm{nadir}},j}}}}{{\partial {{\bar K}_{j}}}} \cdot \frac{{{{\bar K}_{j}}}}{{{{\omega }_{{\rm{nadir}},j}}}}}\\
{\frac{{\partial {{\omega }_{{\rm{nadir}},j}}}}{{\partial {{\bar \tau }_{j}}}} \cdot \frac{{{{\bar \tau }_{j}}}}{{{{\omega }_{{\rm{nadir}},j}}}}}
\end{array}} \right]
\end{equation}
where $\textbf{S}_{{\rm{nadir}},j}$ is the sensitivity matrix of the maximum frequency deviation to the regulation parameters. Since the analytical solutions of \eqref{eq:Smin} are difficult to get, the values are calculated using numerical methods, as shown in Fig.\ref{fig.7}(b). It can be seen that the maximum frequency deviation is far more sensitive to the ENI than to other frequency regulation parameters. Regarding that the same conclusion holds for the maximum RoCoF and the steady-state frequency is constant. The ENI is therefore chosen as the most important parameter that determines the frequency security indicators.

\subsection{Critical Nodal Inertia for Frequency Security}
\subsubsection{Constraint of Maximum RoCoF} Considering the maximum RoCoF constraint, the absolute value of the maximum RoCoF of each node should not exceed the threshold.
\begin{equation}\label{eq:ROCOF requirement}
\left| {{R_{\max ,j}}} \right| \le {R_{{\rm{th}}}},\forall j \in \mathbb{N}
\end{equation}
where $\mathbb{N}$ is the set of key nodes and $R_{\rm{th}}$ is the RoCoF threshold value. Equation \eqref{eq:ROCOF requirement} can be converted to the constraint on the nodal inertia based on \eqref{ROCOF}.
\begin{equation}\label{eq:inertia ROCOF requirement}
{\bar H_{j}} \ge {H_{R}} = \frac{{\left| {\Delta {P_{i2}}} \right|}}{{{R_{{\rm{th}}}}}},\forall j \in \mathbb{N}
\end{equation}
where $H_{R}$ is the critical nodal inertia constrained by the RoCoF requirement.

\subsubsection{Constraint of Maximum Frequency Deviation} Requirement on the Maximum frequency deviation is described as,

\begin{equation}\label{eq:nadir requirement}
{\omega _{{\rm{nadir}},j}} \ge {\omega _{\rm{th}}},\forall j \in \mathbb{N}
\end{equation}
where $\omega_{\rm{th}}$ is the minimum allowable frequency. Denote the relationship between the minimum nodal frequency and the ENI as $\chi_{j}$ based on \eqref{frequency nadir}.
\begin{equation}\label{eq:nadir}
{ \omega _{{\rm{nadir}},j}} \buildrel \Delta \over = {\chi _{j}}\left( {{{\bar H}_{j}}} \right),j \in \mathbb{N}
\end{equation}

Then the constraint of (14) can be converted into \eqref{eq:inertia nadir requirement}.
\begin{equation}\label{eq:inertia nadir requirement}
{\bar H_{j}} \ge {H_{j,{\rm{dev}}}} = \chi _{j}^{ - 1}\left( {{\omega _{\rm{th}}}} \right),\forall j \in \mathbb{N}
\end{equation}
where ${H_{j,{\rm{dev}}}}$ is the critical nodal inertia constrained by the maximum frequency deviation.

\subsubsection{Constraint of Steady-State Frequency} According to the analysis in Section IV-A, the steady-state frequency will finally recover to the nominal value. The steady-state frequency requirement has therefore always been met.

In summary, to satisfy the requirements on the three frequency security indicators, the ENI of all nodes should both met \eqref{eq:inertia ROCOF requirement} and \eqref{eq:inertia nadir requirement}. That is,
\begin{equation}\label{frequency security index}
{\bar H_{j}} \ge {H_{j,\rm{cri}}} = \max \left\{ {{H_{R}},{H_{j,{\rm{dev}}}}} \right\},\forall j \in \mathbb{N}
\end{equation}
where $H_{j\rm{,cri}}$ is the critical nodal inertia to satisfy the frequency security requirements.

\subsection{Proposed Frequency Security Index} 
The node frequency security index of the node $j$ is proposed based on the node inertia, donated as $\kappa _{j}$.
\begin{equation}\label{eq:nodal frequency security index}
{\kappa _{j}} = \frac{{{{\bar H}_{j}} - {H_{j{\rm{,cri}}}}}}{{{H_{j{\rm{,cri}}}}}} \times 100\% , j \in \mathbb{N}
\end{equation}

The system frequency safety shall be determined by the weakest node within the system. Therefore, the frequency security index for the whole power system, donated as $\kappa$, is represented by the minimum nodal frequency security index.
\begin{equation}\label{eq:system frequency security index}
{\kappa} = \min\left\{ {{\kappa _{j}}} \right\}, j \in \mathbb{N}
\end{equation}

The value of the index $\kappa$ could represent the system frequency safety:

\subsubsection{$\kappa>0$} This means that the actual effective inertia of all nodes is larger than the critical inertia. The nodal inertia is significant enough to guarantee the frequency security. Therefore, the system is frequency safe in this case. The larger the index $\kappa_{i}$, the safer the system is. 
\subsubsection{$\kappa=0$} One or all of the frequency dynamics indicators have reached the allowable boundary, the system maintains critical frequency safety.
\subsubsection{$\kappa<0$} The effective nodal inertia of at least one node does not meet the frequency security requirements. This makes the system unsafe.

\subsection{Procedure of Frequency Security Assessment}
\begin{figure}[b]
\centering
\includegraphics[width=80mm]{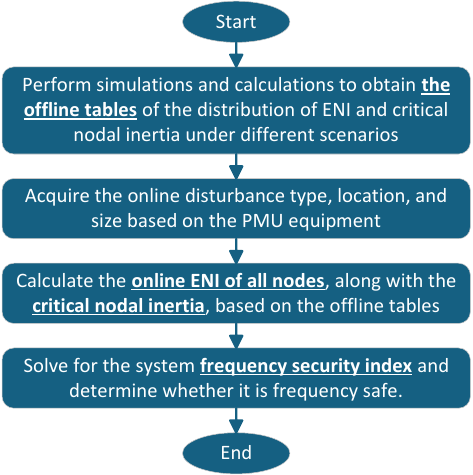}
\caption{Flowchart of the ``offline calculation and online assessment" procedure.}
\label{fig.assess}
\end{figure}

The frequency security assessment process is called ``offline calculation and online assessment", which is shown in Fig.\ref{fig.assess}.

First, numerous simulations should be conducted on the system to obtain the frequency dynamics of all nodes across the system under different disturbances. Or enough measurement data needs to be gained in the real system. Based on the frequency/power data, the ENI and the critical nodal inertia are calculated based on the data-based fitting method proposed in Section III. Then the offline table of ENI and the critical inertia of all nodes under different scenarios are formed.

Second, the type, location, and size of the disturbance are identified based on the online measurement data from the PMU equipment  \cite{mobashsher2023new}.

Third, identify the operating conditions in the offline table that are close to the online operating condition. Then, calculate the ENI and the critical nodal inertia for this online operating condition by applying a weighting based on the Euclidean distance, according to \eqref{eq:online_ENI}.
\begin{equation}
\label{eq:online_ENI}
H_{{\rm{on}}} = \frac{\sum\limits_{r \in C} \frac{H_{{\rm{off}},r}}{\|L_{{\rm{on}}-r}\|_2} }{\sum\limits_{r \in C} \frac{1}{\|L_{{\rm{on}}-r}\|_2}}
\end{equation}
where ${H_{{\rm{on}}}}$ and ${H_{{\rm{off}},r}}$ is the effective/critical nodal inertia of the online case and the offline case, respectively. $C$ is the union of the offline operating condition cases in the offline table that are close to the online operating case. ${\|L_{{\rm{on}}-r}\|_2}$ is the Euclidean distance between the online case and the offline case nearby.

Finally, the system frequency security index is obtained based on \eqref{eq:system frequency security index}. The system is then evaluated to determine whether it is frequency safe.

The advantages of the proposed methods are listed as below.
\begin{enumerate}
    \renewcommand{\labelenumi}{(\arabic{enumi})}
    \item The nodal inertia-based frequency security index features a simple calculation process, which is highly suitable for online frequency security assessment in engineering practice.
    \item The method is applicable to any disturbance condition, including unforeseen events not covered in the offline dataset, with strong generalization and robustness.
    \item The weak nodes that dominate system frequency security can be identified. This could provide quantitative guidance for generation configuration and optimization.
\end{enumerate}

\section{Case Studies}
\begin{figure}[h]
\centering
\includegraphics[width=88mm]{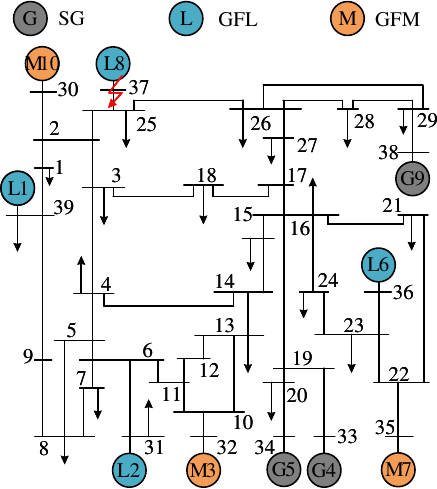}
\caption{Modified IEEE 39-bus system.}
\label{fig.6-0}
\end{figure}

Simulations are conducted in the modified IEEE 39-bus system, which is depicted in Fig.\ref{fig.6-0}. Parameters of the generations are shown in TABLE \ref{tab:generator_params}.

\begin{table}[h]
  \centering
  \caption{Parameters of all generators}
  \label{tab:generator_params}
  \begin{tabular}{lccccc}
    \toprule
    Parameter & L1 & L2 & M3 & G4 & G5 \\
    \midrule
    \(P\) (MW) & 501.2 & 250.5 & 611 & 1000 & 510 \\
    %\midrule
    \(T\) (s) & -- & -- & 0.5 & 1 & \begin{tabular}{@{}c@{}}5 (case I)\\3 (case II)\end{tabular}  \\
    %\midrule
    \(D\) (pu/pu) & -- & -- & 20 & 6 & 6 \\
    %\midrule
    {$K_{\text{pri}}$ (pu/pu)} & -- & -- & -- & 20 & 20 \\
    %\midrule
    {$\tau$ (s)} & -- & -- & -- & 1 & 1 \\
    \midrule
    Parameter & L6 & M7 & L8 & G9 & M10 \\
    \midrule
    \(P\) (MW) & 650.6 & 576 & 540.7 & 830 & 561 \\
    %\midrule
    \(T\) (s) & -- & 0.5 & -- & \begin{tabular}{@{}c@{}}10 (case I)\\6 (case II)\end{tabular} & 0.5 \\
    %\midrule
    \(D\) (pu/pu) & -- & 20 & -- & 6 & 20 \\
    %\midrule
    {$K_{\text{pri}}$ (pu/pu)} & -- & -- & -- & 20 & -- \\
    %\midrule
    {$\tau$ (s)} & -- & -- & -- & 1 & -- \\
    \bottomrule
  \end{tabular}
\end{table}

\begin{figure}[t]
\centering
\includegraphics[width=88mm]{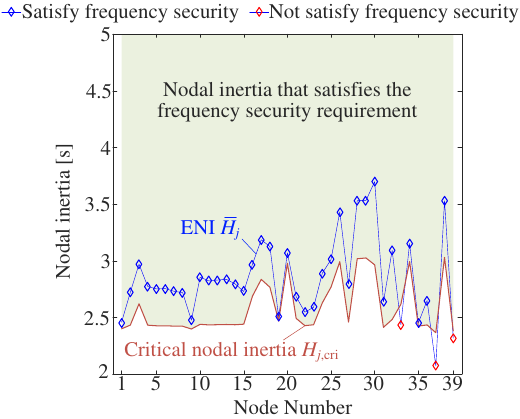}
\caption{The critical nodal inertia and ENI results of the online case II.}
\label{fig.8}
\end{figure}

\subsection{Critical Nodal Inertia and ENI Results} 
Take case II as an example, the critical nodal inertia constraint by the frequency security is plotted with the red line in Fig.\ref{fig.8}. It can be seen that the critical nodal inertia ranges from 2.4 s to 3 s. The allowable inertia range to satisfy the frequency security requirement is drawn with green. Similarly, the online ENIs of the 39 nodes are also drawn in Fig.\ref{fig.8} with the blue line. It can be seen that the inertia of the node ${\rm{N}}_{37}$ is the lowest of about 2.1 s, and the inertia of the node ${\rm{N}}_{30}$ is the highest of about 3.6 s. The difference in inertia between different nodes reaches 1.5 s, which is approximately 70 $\%$ of the nodal inertia of the node ${\rm{N}}_{37}$. This indicates that there are significant differences in the inertia levels across different regions. Comparing the ENI and the critical nodal inertia, whether the frequency security is satisfied at this node is concluded. The ENIs of the nodes ${\rm{N}}_{33}$, ${\rm{N}}_{37}$, and ${\rm{N}}_{39}$ are smaller than the corresponding critical inertia. Therefore, frequency security problems would occur in these nodes/areas.

\subsection{Frequency Security Assessment Results} 
\begin{figure}[t]
\centering
\includegraphics[width=88mm]{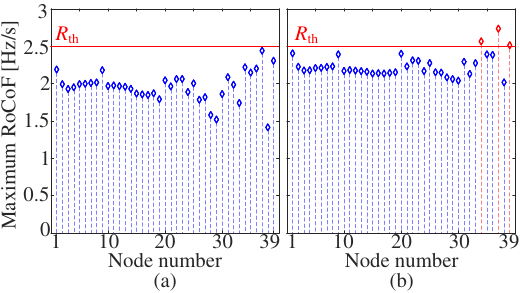}
\caption{Maximum RoCoF under (a) case I and (b) case II.}
\label{fig.9}
\end{figure}

\begin{figure}[t]
\centering
\includegraphics[width=88mm]{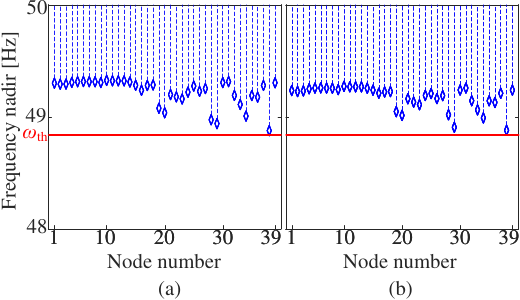}
\caption{The maximum frequency deviation under (a) case I and (b) case II.}
\label{fig.10}
\end{figure}

\begin{figure}[t]
\centering
\includegraphics[width=88mm]{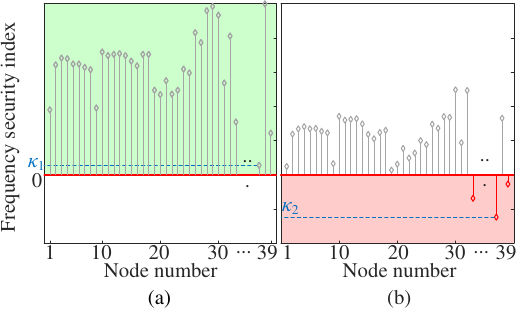}
\caption{SFSA results of (a) cases I and (b) case II.}
\label{fig.11}
\end{figure}

The maximum RoCoFs of all 39 nodes in case I are plotted with blue diamonds in Fig.\ref{fig.9} (a). The maximum frequency deviation of case I is drawn with blue diamonds in Fig.\ref{fig.10} (a). These results show that the frequency security indicators of all nodes are within the allowable range, indicating the system is frequency safe in case I. The frequency security assessment results using the proposed method in this paper is displayed in Fig.\ref{fig.11} (a). The frequency security index ${\kappa}$ is larger than zero. Based on the analysis in Section IV-D, the system is evaluated to be frequency safe with the proposed method, which is consistent with the simulation results of Fig.\ref{fig.9} (a) and Fig.\ref{fig.10} (a).

Compared to case I, the inertia of the G5 is reduced from 5 s to 3 s, and the inertia of the G9 is decreased from 10 s to 6 s in case II. The maximum RoCoF and frequency deviation results of case II are depicted in Fig.\ref{fig.9} (b) and Fig.\ref{fig.10} (b), respectively. These results suggest that the RoCoF is obeyed in node ${\rm{N}}_{33}$, ${\rm{N}}_{37}$,and ${\rm{N}}_{39}$, even though the maximum frequency deviation of all nodes met the requirements. Therefore, frequency security problems would occur in this case. Fig.\ref{fig.11} (b) shows the assessment result using the proposed method. The frequency security index of the whole system is found to be smaller than zero, indicating that the system is not able to maintain frequency security. Furthermore, this method identified the weak area which is near nodes ${\rm{N}}_{33}$, ${\rm{N}}_{37}$, and ${\rm{N}}_{39}$. To prevent from system frequency security problems, the frequency stability of these nodes, especially the node ${\rm{N}}_{37}$, should be improved using inertia-boosting strategies such as adding energy storage and increasing the inertia control parameters of inverters \cite{he2026inertiamatchingprincipleimproving}.

\section{Conclusions}
As low-inertia converter-controlled RESs replaces traditional SGs featuring large rotor inertia, frequency security of power systems with high penetration of RESs is facing a serious challenge. The frequency is more susceptible to disturbances, and the spatial-temporal heterogeneity of regional frequency responses is further exacerbated, making it more difficult to accurately assess the frequency security. To address this issues, an online frequency security assessment method considering the spatio-temporal frequency distribution is proposed in this paper, with nodal inertia as the core characterization parameter. A simplified generic ENF model is built with constant effective parameters that dominate the nodal frequency characteristics, ensuring both model simplicity and the retention of essential frequency dynamics. Based on the model, the analytical expressions for the nodal frequency trajectory and the frequency security indicators are derived. The ENI is then discovered to have the most important influence on the frequency security. Consequently, the critical nodal inertia to meet the frequency security requirement is then calculated. An index integrating the ENI and the nodal critical inertia is proposed for assessing frequency security, which considers the frequency distribution and outperforms traditional average methods that ignore spatial heterogeneity. When applied to online scenarios, an interpolation method is employed to quickly determine the ENI and critical nodal inertia based on offline lookup tables, thereby developing an offline calculation and online assessment" method for SFSA.

\appendix[Analytical Expression of Generic ENF Model and Maximum Nodal Frequency Deviation]
\renewcommand{\theequation}{\Alph{section}.\arabic{equation}}
% \Alph{section} = A,B,C...
% 最终编号：A.1, A.2, B.1 ...
Denote $\Delta {\bar \omega _j}$ as the difference between the effective frequency of the node $j$ and the nominal frequency. That is,
\begin{equation}\label{eq:frequency_diff}
\Delta {\bar \omega _j} = {\bar \omega _j} - {\omega _{\rm{0}}}
\end{equation}

Based on \eqref{eq:simple nodal frequency}, the closed-loop transfer function from the power disturbance to the frequency difference is given.
\begin{equation}
\label{eq:frequency_transfer_per}
\frac{G_{\Delta \omega}(s)}{G_{\Delta P}(s)} = \frac{\bar \tau_{j} s+1}{2\bar \tau_{j} \bar H_{j}s^2+(\bar \tau_{j} \bar D_{j}+2\bar H_{j})s+\bar D_{j}+\bar K_{j}}
\end{equation}

\noindent\textit{I. Permanent Power Disturbance}

The Laplace function of the power disturbance is $G_{\Delta P}(s) = {\Delta P_i}/{s}$. Therefore, the time-domain solution of the frequency difference is derived as,
\begin{equation}
\label{eq:frequency_laplace_per}
\Delta \bar \omega_j(t) = {\cal L}^{-1}\left\{ \frac{G_{\Delta \omega}(s)}{G_{\Delta P}(s)}\frac{\Delta P_i}{s}\right\}
\end{equation}

The ENF dynamics are then solved as,
\begin{equation}
\label{eq:frequency_solut_per}
{\bar \omega _j}(t) = \omega_0 + \frac{{\Delta {P_i}}}{{\bar D_{j} + \bar K_{j}}} \cdot \left[ {\begin{array}{*{20}{c}}
1 - {e^{ - \lambda \left( {t - {t_{\rm{f}}}} \right)}} \cdot ({\cos \left( {{\omega _d}\left( {t - {t_{\rm{f}}}} \right)} \right)} \\
{- A_1\sin \left( {{\omega _d}\left( {t - {t_{\rm{f}}}} \right)} \right)} )
\end{array}} \right]\\
\end{equation}
where $\lambda, \omega_d$, and $A_1$ are all constants.

\begin{equation}
\label{eq:varible_expla_per}
\begin{aligned}
\begin{array}{l}
\lambda = \frac{\bar \tau_{j} \bar D_{j} + 2\bar H_{j}}{4\bar \tau_{j} \bar H_{j}} \\
\omega_d = \frac{\sqrt{8\bar \tau_{j} \bar H_{j}(\bar D_{j}+\bar K_{j}) - (\bar \tau_{j} \bar D_{j} + 2\bar H_{j})^2}}{4\bar \tau_{j} \bar H_{j}}\\
A_1 = \frac{2\bar \tau_{j} \bar K_{j} - \bar \tau_{j} \bar D_{j} - 6\bar H_{j}}{\sqrt{8\bar \tau_{j} \bar H_{j}(\bar D_{j}+\bar K_{j}) - (\bar \tau_{j} \bar D_{j} + 2\bar H_{j})^2}}
\end{array}
\end{aligned}
\end{equation}
% \begin{subequations}
% \label{eq:varible_expla_per}
% \begin{align}
% \lambda &= \frac{\bar \tau_{j} \bar D_{j} + 2\bar H_{j}}{4\bar \tau_{j} \bar H_{j}} \\
% \omega_d &= \frac{\sqrt{8\bar \tau_{j} \bar H_{j}(\bar D_{j}+\bar K_{j}) - (\bar \tau_{j} \bar D_{j} + 2\bar H_{j})^2}}{4\bar \tau_{j} \bar H_{j}}\\
% A_1 &= \frac{2\bar \tau_{j} \bar K_{j} - \bar \tau_{j} \bar D_{j} - 6\bar H_{j}}{\sqrt{8\bar \tau_{j} \bar H_{j}(\bar D_{j}+\bar K_{j}) - (\bar \tau_{j} \bar D_{j} + 2\bar H_{j})^2}}
% \end{align}
% \end{subequations}

\noindent\textit{II. Temporary Power Disturbance}

The nodal frequency dynamics can be divided into four periods: pre-fault, fault-on, fault-recovery, and post-fault phases. 

\noindent \textit{(1) Pre-fault phase ($0\le t\le t_{\rm{f}}$)}. In this stage, the (effective) nodal frequency remains at the nominal frequency.
\begin{equation}
\label{eq:pre_fault}
\bar \omega_j(t) = \omega_0, 0\le t\le t_{\rm{f}}
\end{equation}

\noindent \textit{(2) Fault-on phase ($t_{\rm{f}}<t\le t_{\rm{c}}$)}. Based on the analysis in Section III-A, the power deficit is a constant of $\Delta P_{i2}$ in this stage. The ENF dynamics have the same form as \eqref{eq:varible_expla_per}.
\begin{equation}
\label{eq:frequency_solut_per2}
{\bar \omega _j}(t) = \omega_0 + K_2\cdot\left[ {\begin{array}{*{20}{c}}
{1 - {e^{ - \lambda \left( {t - {t_{\rm{f}}}} \right)}} \cdot }\\
{\left( {\cos \left( {{\omega _d}\left( {t - {t_{\rm{f}}}} \right)} \right) - A_1\sin \left( {{\omega _d}\left( {t - {t_{\rm{f}}}} \right)} \right)} \right)}
\end{array}} \right]\\
\end{equation}
where $K_2 = {{\Delta {P_{i2}}}}/({{\bar D_{j} + \bar K_{j}}})$.
The time derivative of the frequency dynamics \eqref{eq:frequency_solut_per2} in this stage is calculated as,
\begin{equation}
\label{eq:frequency_deriva2}
{\bar \omega '_j}(t) = {K_2}{e^{ - \lambda (t - {t_{\rm{f}}})}} \cdot \left[ {\begin{array}{*{20}{c}}
{\left( {\lambda  + {A_1}{\omega _d}} \right)\cos \left( {{\omega _d}(t - {t_{\rm{f}}})} \right) + }\\
{\left( {{\omega _d} - \lambda {A_1}} \right)\sin \left( {{\omega _d}(t - {t_{\rm{f}}})} \right)}
\end{array}} \right]
\end{equation}

Find the time $t^*_2$ at which the frequency derivative is zero. 
\begin{equation}
\label{nadir2}
{t^*_{\rm{2}} = t_{\rm{f}} + \frac{1}{\omega_d} \left( k\pi - \arctan\frac{\lambda + A_1 \omega_d}{\lambda A_1 - \omega_d} \right), k \in R_N}
\end{equation}
where $R_N$ is the set of integers. If there exists an minimum integer $k$ that satisfies $t^*_{\rm{2}}\in\left[t_{\rm{f}},t_{\rm{c}} \right]$, $t_{\rm{m2}} = t^*_2$. Otherwise, if there is no $k$ satisfying $t^*_{\rm{2}}\in\left[t_{\rm{f}},t_{\rm{c}} \right]$,  $t_{\rm{m2}} = \arg \min \left\{ {{{\bar \omega }_j}({t_{\rm{f}}}),{{\bar \omega }_j}({t_{\rm{c}}})} \right\}$.

\noindent \textit{(3) Fault-recovery phase ($t_{\rm{c}}<t\le t_{\rm{r}}$)}. According to the analysis in Section III-A, The power deficit is $\Delta P_{i3} = (r_{\rm{p}}t – 1)P_{\rm{load0}}$ in this stage. The initial states of this stage (${{\bar \omega }_j}({t_{\rm{c}}})$ and ${{\bar \omega }'_j}({t_{\rm{c}}})$) can be calculated based on \eqref{eq:frequency_solut_per2}. Then the time-domain solution of the ENF is derived based on the Laplace transform method.
\begin{equation}
\label{eq:frequency_solut_per3}
{{\bar \omega }_j}\left( t \right) = \left[ {\begin{array}{*{20}{c}}
{{\omega _0} + {C_1}\left( {t - {t_{\rm{c}}}} \right) + {C_2} + {e^{ - \lambda \left(t-t_{\rm{c}}\right)}} \cdot }\\
{\left( {{A_{\cos }}\cos \left( {{\omega _d}\left( {t - {t_{\rm{c}}}} \right)} \right) + {A_{\sin }}\sin \left( {{\omega _d}\left( {t - {t_{\rm{c}}}} \right)} \right)} \right)}
\end{array}} \right]
\end{equation}
where $C_1, C_2, A_{\rm{cos}}$, and $A_{\rm{sin}}$ are all constants.

\begin{equation}
\label{eq:varible_expla_per3}
\begin{aligned}
\begin{array}{l}
{C_1} = \frac{{r_{\rm{p}}{P_{{\rm{load0}}}}}}{{\bar D_{j} + \bar K_{j}}}\\
{C_2} = {P_{{\rm{load0}}}} \cdot \frac{{r_{\rm{p}}\bar \tau \bar K_{j} - \left( {\bar D_{j} + \bar K_{j}} \right) - 2r_{\rm{p}}\bar H_{j}}}{{{{\left( {\bar D_{j} + \bar K_{j}} \right)}^2}}}\\
{A_{\cos }} = {{\bar \omega }_j}\left( {{t_{\rm{c}}}} \right) - \frac{{{P_{{\rm{load0}}}}\left( {r_{\rm{p}}\bar \tau_{j} \bar K_{j} - \bar D_{j} - \bar K_{j} - 2r_{\rm{p}}\bar H_{j}} \right)}}{{{{\left( {\bar D_{j} + \bar K_{j}} \right)}^2}}}\\
{A_{\sin }} = \frac{{2\bar \tau_{j} \bar H_{j}{\bar \omega'_j}\left( {{t_{\rm{c}}}} \right) + \frac{{\left({{\bar \omega }_j}\left( {{t_{\rm{c}}}} \right)-C_2\right)\left( {\bar \tau_{j} \bar D_{j} + 2\bar H_{j}} \right)}}{2} - \frac{{2r_{\rm{p}}\bar \tau_{j} \bar H_{j}{P_{{\rm{load0}}}}}}{{\bar D_{j} + \bar K_{j}}}}}{{2\bar \tau_{j} \bar H_{j}{\omega _d}}}
\end{array}
\end{aligned}
\end{equation}

Similarly, the derivative of the frequency in this stage is calculated as,
\begin{equation}
\label{eq:nadir3}
{{\bar \omega '}_j}({t}) =  \left[ {\begin{array}{*{20}{c}}
{{C_1} +{e^{ - \lambda ({t} - {t_{\rm{c}}})}} \cdot }\\
{\left[ {\begin{array}{*{20}{c}}
{\left( {{\omega _d}{A_{\sin }} - \lambda {A_{\cos }}} \right)\cos \left( {{\omega _d}({t} - {t_{\rm{c}}})} \right)}-\\
{ \left( {{\omega _d}{A_{\cos }} + \lambda {A_{\sin }}} \right)\sin \left( {{\omega _d}({t} - {t_{\rm{c}}})} \right)}
\end{array}} \right]}
\end{array}} \right]
\end{equation}

Setting equation \eqref{eq:nadir3} to zero gives
\begin{equation}
\label{eq:nadir3-2}
e^{-\lambda \tau} \cos(\omega_d \tau+\alpha) = \gamma
\end{equation}
where
\begin{equation}
\label{eq:nadir3-2-2}
\begin{aligned}
\tau &= t-t_{\rm{c}}\\
\alpha &= \arctan(\frac{\omega_dA_{\cos}+\lambda A_{\sin}}{\omega_d A_{\sin}-\lambda A_{\cos}})\\
\gamma &= -\frac{C_1}{\sqrt{(\omega_d A_{\sin}-\lambda A_{\cos})^2+(\omega_d A_{\cos}+\lambda A_{\sin})^2}}
\end{aligned}
\end{equation}

For simplification, a second-order Taylor expansion around $\tau = 0$ is performed to equation \eqref{eq:nadir3-2}. That is,
\begin{equation}
\label{eq:Taylor expansion}
\begin{aligned}
e^{-\lambda \tau} &\approx 1-\lambda \tau+\frac{1}{2}\lambda^2\tau^2\\
\cos(\omega_d \tau+\alpha) &\approx \cos \alpha - \omega_d \sin \alpha \cdot \tau - \frac{1}{2}\omega^2_d \cos \alpha \cdot \tau^2\\
\end{aligned}
\end{equation}

When ignoring the high-order terms, \eqref{eq:nadir3-2} is simplified.
\begin{equation}
\label{solution_simple}
\begin{aligned}
&\underbrace{\cos\alpha - \gamma}_{E_0} - \underbrace{\left( \lambda \cos\alpha + \omega_d \sin\alpha \right)}_{E_1}\tau \\
&+ \underbrace{\frac{1}{2}\left( \left( \lambda^2 - \omega_d^2 \right) \cos\alpha - 2\lambda\omega_d \sin\alpha \right)}_{E_2}\tau^2 = 0
\end{aligned}
\end{equation}

Then the solution is calculated as,
\begin{equation}
\tau^* = \frac{E_1 - \sqrt{E_1^2 - 4 E_2 E_0}}{2 E_2},
\quad t^*_{\rm{3}} = t_{\rm{c}} + \tau^*.
\end{equation}

If $t^*_{\rm{3}}\in\left[t_{\rm{c}},t_{\rm{r}} \right]$, $t_{\rm{m3}} = t^*_3$. Otherwise, if $t^*_{\rm{3}}\notin\left[t_{\rm{c}},t_{\rm{r}} \right]$,  $t_{\rm{m3}} = \arg \min \left\{ {{{\bar \omega }_j}({t_{\rm{c}}}),{{\bar \omega }_j}({t_{\rm{r}}})} \right\}$.

\noindent \textit{(4) Post-fault phase ($t> t_{\rm{r}}$)}. After the time of $t_{\rm{c}}$, all the loads and generators recover to the pre-fault states, then the frequency dynamics have the same form with \eqref{eq:frequency_solut_per2}. The initial states of this stage are ${{\bar \omega }_j}({t_{\rm{r}}})$ and ${{\bar \omega }'_j}({t_{\rm{r}}})$. The frequency dynamics are derived using the Laplace transform method.
\begin{equation}
\label{eq:frequency_solut_per4}
{{\bar \omega }_j}(t) = {\omega _0} + {e^{ - \lambda (t-t_{\rm{r}})}}\cdot\left[ {\begin{array}{*{20}{c}}
{{{\bar \omega }_j}\left( {{t_{\rm{r}}}} \right)\cos ({\omega _d}(t - {t_r})) + }\\
{\frac{{{{\bar \omega '}_j}({t_{\rm{r}}}) + \lambda {{\bar \omega }_j}\left( {{t_{\rm{r}}}} \right)}}{{{\omega _d}}}\sin ({\omega _d}(t - {t_{\rm{r}}}))}
\end{array}} \right]
\end{equation}

The time deviation of the frequency is,
\begin{equation}
{\bar \omega '_j}(t) = {e^{ - \lambda (t-t_{\rm{r}})}} \cdot \left[ {\begin{array}{*{20}{c}}
{{{\bar \omega }'_{j}}({t_{\rm{r}}})\cos ({\omega _d}(t - {t_{\rm{r}}})) - }\\
{\frac{{\lambda {{\bar \omega }'_{j}}({t_{\rm{r}}}) + (\omega _d^2 + {\lambda ^2}){{\bar \omega }_j}({t_{\rm{r}}})}}{{{\omega _d}}}\sin ({\omega _d}(t - {t_{\rm{r}}}))}
\end{array}} \right]
\label{eq:frequency_derivative}
\end{equation}

Solution of this phase has the same form as  
\begin{equation}
t^*_{\rm{4}} = t_{\rm{r}} + \frac{1}{\omega_d} \left[ \arctan\left( \frac{\omega_d \bar{\omega}_j'(t_{\rm{r}})}{\lambda \bar{\omega}_j'(t_{\rm{r}}) + (\omega^2_d+\lambda^2) \bar{\omega}_j(t_{\rm{r}})} \right) + k\pi \right]
\label{eq:extreme_time}
\end{equation}

Let $k^*_4$ be the minimum integer satisfying $t^*_4 > t_{\rm{r}}$. Then $t_{\rm{m4}}$ is equal to $t^*_{\rm{4}}$ at $k = k^*_4$ based on \eqref{eq:extreme_time}.

\bibliographystyle{ieeetr}
\bibliography{ref.bib}

\end{document}